\documentclass[%
 reprint,
 amsmath,amssymb,
 aps,
]{revtex4-2}

\usepackage{graphicx}
\usepackage{dcolumn}
\usepackage{bm}

\begin{document}


\title{Hybrid multi-bend achromat lattice with sextupole cancellation across straight section}
\author{Jiajie Tan}
\author{Jianhao Xu}%
\author{Penghui Yang}%
\author{Zhenghe Bai}%
 \email{baizhe@ustc.edu.cn} 
\author{Lin Wang}%
 \email{wanglin@ustc.edu.cn} 
\affiliation{%
 National Synchrotron Radiation Laboratory, University of Science and Technology of China, Hefei 230029, China
}%

\date{\today}

\begin{abstract}

The hybrid multi-bend achromat (HMBA) lattice concept is adopted in some diffraction-limited storage ring designs, which can permit relatively large on-momentum dynamic aperture and relatively weak sextupoles. In a typical HMBA lattice, the main arc section is constrained by the transverse phase advances making –$\it{I}$ transformation for sextupole cancellation. In this paper, a new HMBA lattice concept with sextupole cancellation across straight section is proposed, where –$\it{I}$ is made between adjacent dispersion bumps of two lattice cells. This makes the main arc section free of the phase advance constraint, and as a result, the number of bending magnets (bends) in the lattice cell and the cell tunes can be easily changed, thus providing more choices for lattice design. To achieve the large phase advances required for –$\it{I}$ in this new concept, split bend is used as the matching bend, which is a bend split into two pieces with a quadrupole in between. The split bend also serves to reduce the emittance, and the large phase advances also give low beta functions in the straight section enhancing the insertion device brightness. Besides, for a given emittance goal, this new HMBA lattice can have less bends than the typical HMBA lattice due to stronger focusing in bend unit cells, which is beneficial for saving space and suppressing intra-beam scattering effect. Two lattices are given as examples to demonstrate this new concept and show its linear and nonlinear properties, and further extension is also discussed.

\end{abstract}

\maketitle

\section{Introduction}

We can relate the natural emittance of an electron storage ring to several parameters as
\begin{equation}
\varepsilon_{x}={F}\frac{E^{2}}{J_{x}N_{b}^{3}},
\end{equation}
where $F$ is a ring lattice dependent factor, $E$ is the electron energy, $J_x$ is the horizontal damping partition number, and $N_b$ is the number of bending magnets (bends) in the ring. Due to the inverse third-power dependence of $\varepsilon_{x}$ on $N_b$, multi-bend achromat (MBA) lattices \cite{ref-1-MBA} are used for designing diffraction-limited storage rings (DLSRs). Among them, the hybrid MBA (HMBA) lattice concept \cite{ref-2-HMBA}, developed by ESRF-EBS, has been adopted in some DLSR designs for its relatively large on-momentum dynamic aperture (DA) and relatively weak sextupoles. This is attributed to the creation of a pair of dispersion bumps on both sides of the lattice arc section, separated by a –$\it{I}$ transformation with horizontal and vertical phase advances of (3$\pi$, $\pi$), which can make a high efficiency of chromaticity correction and very effective sextupole cancellation. However, the constraint of phase advances also limits the flexibility of HMBA lattice. The number of bends in the lattice cell can not be easily changed, and it is also difficult to reduce the lattice factor $F$.

Some variants of HMBA lattice have been proposed to pursue better performances and different requirements for DLSRs. The H7BA lattice of APS-U \cite{ref-3-APSU} introduced reverse bends (RBs) \cite{ref-4-antibend} to reduce the emittance; the H6BA lattices of Diamond-II \cite{ref-5-Diamond} and HALF \cite{ref-6-HALF} have an additional straight section in the middle part replacing the central bend to accommodate more insertion devices (IDs). In these lattices, the phase advances between two pumps for –$\it{I}$ remain unchanged. Choosing other phase advances for –$\it{I}$ can give different number of bends, such as (5$\pi$, $\pi$) for an H10BA lattice \cite{ref-7-H10BA} and ($\pi$, $\pi$) for the PETRA IV lattice \cite{ref-8-PETRA-H6BA}. However, such a choice is discontinuous. The number of bends in the lattice cell is restricted to a large extent, and so are the cell tunes. Variable number of bends and adjustable cell tunes are beneficial for lattice design, which can for example reduce the emittance and better satisfy other needs.

Compared to some conventional MBA lattices \cite{ref-9-MBA-lattice1,ref-10-MBA-lattice2,ref-11-MBA-lattice3}, HMBA lattices need more bends to reach the same emittance goal \cite{ref-12} due to relatively weak focusing in their bend unit cells (corresponding to relatively large lattice factor $F$), which will make weaker dipole fields and thus longer damping times. For example, the horizontal tune of the H7BA lattice cell is obviously smaller than that of the SLS-2 7BA lattice cell. Longer damping times can cause more serious intra-beam scattering (IBS) effect. This has been reflected in recently-designed medium-energy HMBA lattices with emittances of only tens of pm$\cdot$rad \cite{ref-13-SSRL-X,ref-14-korean}, where higher energies of 3.5$\sim$4.0 GeV are chosen and also damping wigglers (DWs) are employed to significantly enhance radiation damping for suppressing IBS-induced emittance increase and reducing emittance. Therefore, if the HMBA lattice factor $F$ is reduced by increasing the focusing of bend unit cells, less bends can be needed for a given emittance goal, which is good for suppressing IBS and saving space. However, this is also limited by the phase advance constraint in the arc section.

Moving the phase advance constraint from the main arc section to the both sides of the lattice cell would solve the problems above. In this paper we propose a new HMBA lattice concept with sextupole cancellation across straight section, in which –$\it{I}$ is made between adjacent bumps of two cells. Due to that the main arc section is no longer constrained by the phase advances, it is easy to change the number of bends and the cell tunes. However, to realize this concept, we need to increase the phase advances of both sides of the cell to about (3$\pi$, $\pi$) for making –$\it{I}$, which are too large in the normal magnet layout. Inspired by the studies in Refs. \cite{ref-15,ref-16-splitbend} we use split bend as the matching bend to increase the phase advances, and also the beta functions in the straight section are lowered. The former can also reduce the emittance, and the latter enhances the ID brightness. Besides, the relatively strong focusing in the split bends can also make the focusing in other bends stronger due to the need for optics matching, thus reducing the lattice factor $F$.

The content of this paper is organized as follows. In Sec. II, we describe the new HMBA lattice concept, and then some related properties of split bend cell employed in the concept are studied using simplified lattice model. In Sec. III, two HMBA lattices are presented as examples to illustrate the lattice concept and also to show the lattice properties. In Sec. IV, possible lattice extension based on the sextupole cancellation of the concept is discussed. Finally, we summarize the conclusion in Sec. IV.

\section{Lattice concept}

\begin{figure}
\includegraphics[width=0.95\linewidth]{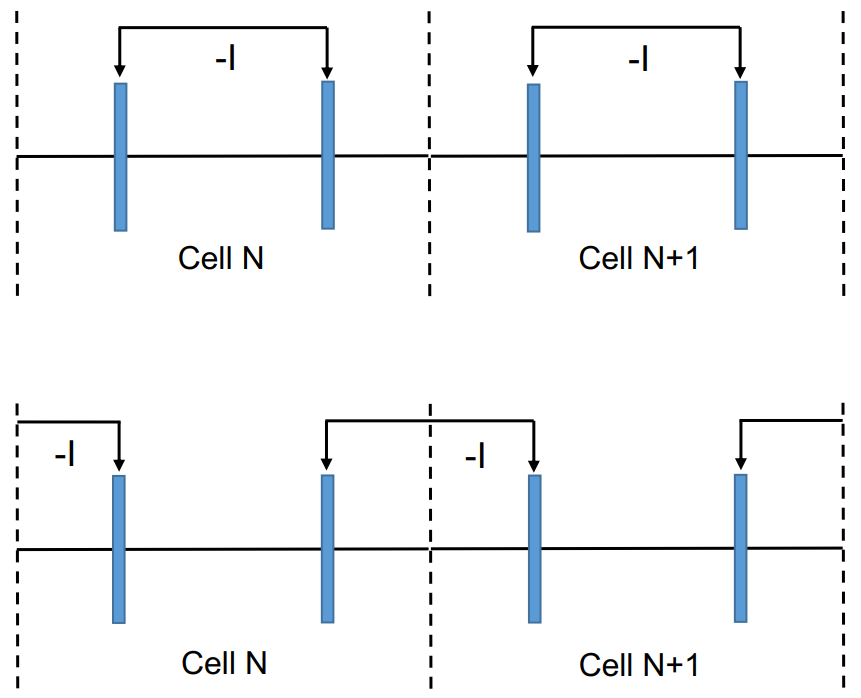}
\caption{\label{fig:f1_-I} Schematic of –$\it{I}$ transformation between two dispersion bumps for the typical HMBA lattice (upper) and new HMBA lattice (lower).}
\end{figure}

The upper part of Fig.~\ref{fig:f1_-I} shows the schematic of –$\it{I}$ transformation made between two dispersion bumps of the typical HMBA lattice cell. The number of bends in the lattice cell and the cell tunes are largely dependent on the inner part of the arc section, which is however constrained by the phase advances for –$\it{I}$. To have better flexibility, here we propose a new HMBA lattice concept with sextupole cancellation across straight section. As shown in the lower part of Fig.~\ref{fig:f1_-I}, –$\it{I}$ is made between adjacent dispersion bumps of two lattice cells, which will make the inner part of the arc section free of the phase advance constraint. To do so, we need to get –$\it{I}$ with the matching sections and straight section, which is however difficult to realize with a normal magnet layout. In most HMBA lattice designs \cite{ref-2-HMBA,ref-3-APSU,ref-5-Diamond,ref-6-HALF}, the horizontal and vertical phase advances of the matching sections and straight section, denoted as ($\Delta\phi_{x}$, $\Delta\phi_{y}$), are about (0.9×2$\pi$, 0.35×2$\pi$), which are far from the required (1.5×2$\pi$, 0.5×2$\pi$), especially in the horizontal plane. SOLEIL designed an HMBA lattice \cite{ref-17-SOLEIL} with very small horizontal beta function in the straight section, which has a larger $\Delta\phi_{x}$ of about 1.2×2$\pi$. But it is still not enough to satisfy the required value for –$\it{I}$.

\begin{figure}
\includegraphics[width=0.95\linewidth]{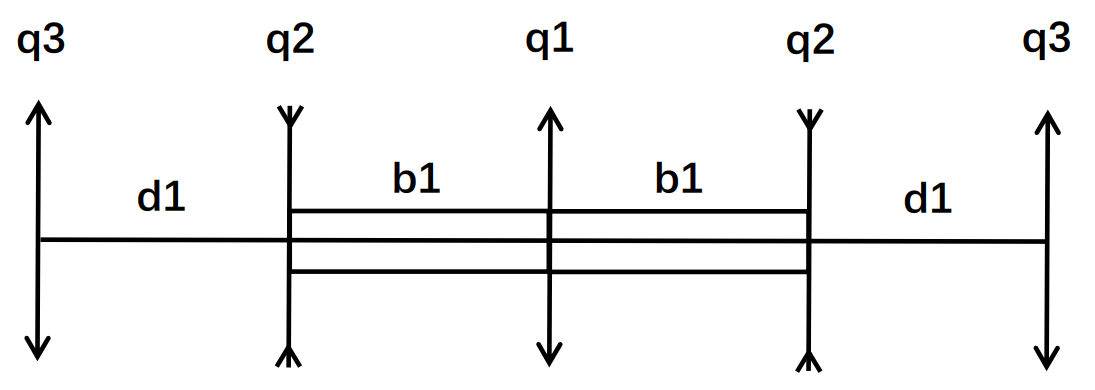}
\caption{\label{fig:f2_SB_TME} Layout of the simplified unit cell model with split bend. In the model, $q1$ and $q3$ are horizontally focusing quadrupoles and $q2$ is defocusing quadrupole.}
\end{figure}

\begin{figure}
\includegraphics[width=0.95\linewidth]{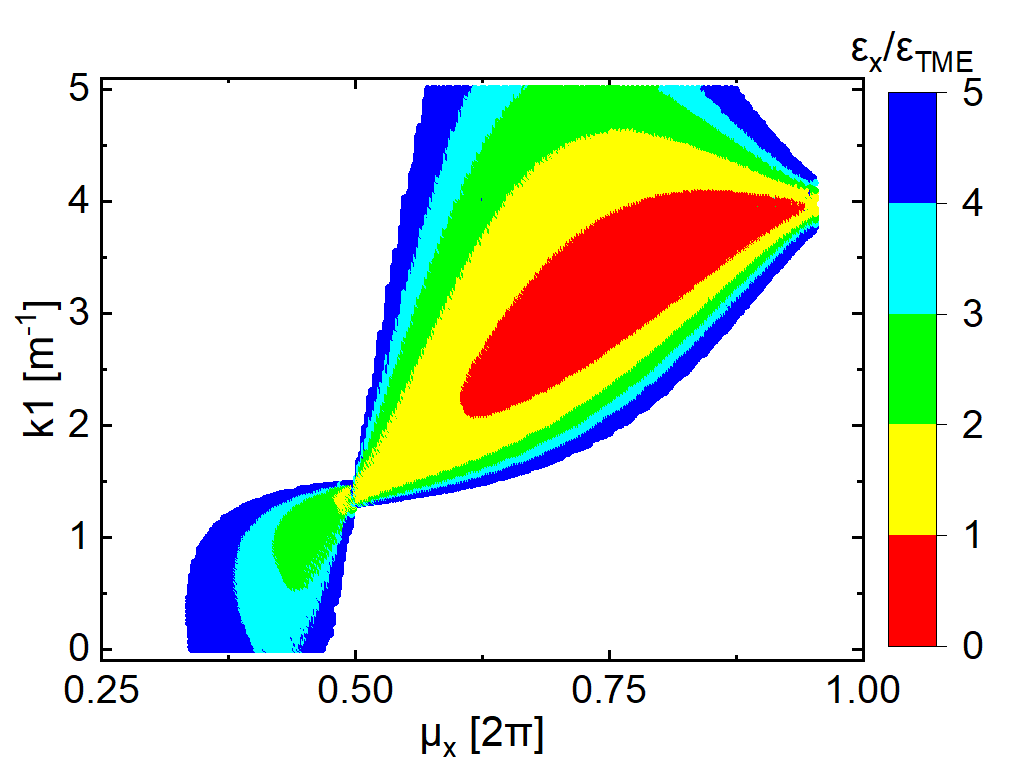}
\caption{\label{fig:f3_SB_scan_1} Cell horizontal phase advance $\mu_{x}$ vs. quadrupole strength $k1$ for scaned solutions, with colour denoting the emittance ratio $\varepsilon_{x}/\varepsilon_{TME}$.}
\end{figure}

\begin{figure*}
    \begin{minipage}[t]{0.48\linewidth}
		\centering
		\includegraphics{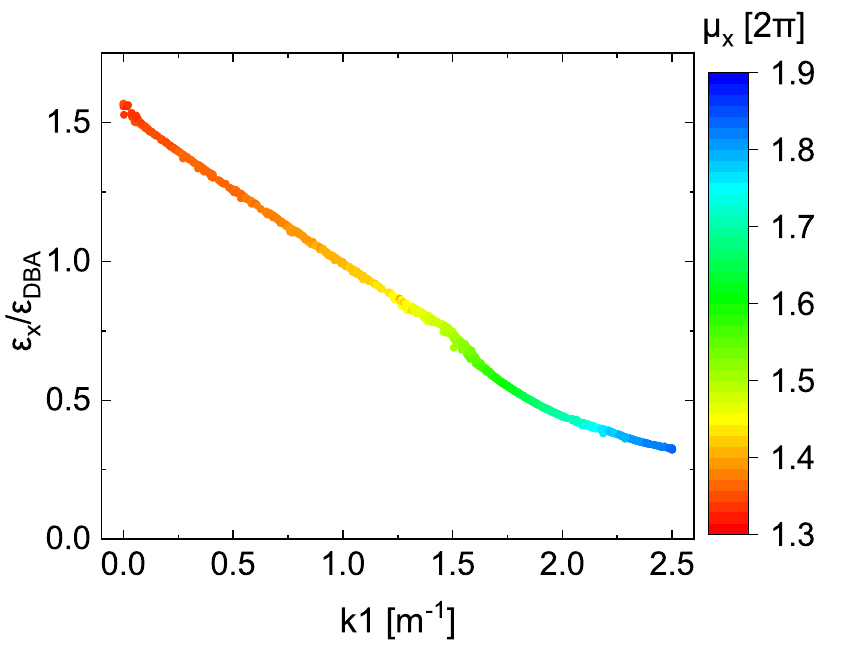}
    \end{minipage}
    \begin{minipage}[t]{0.48\linewidth}
		\centering
		\includegraphics{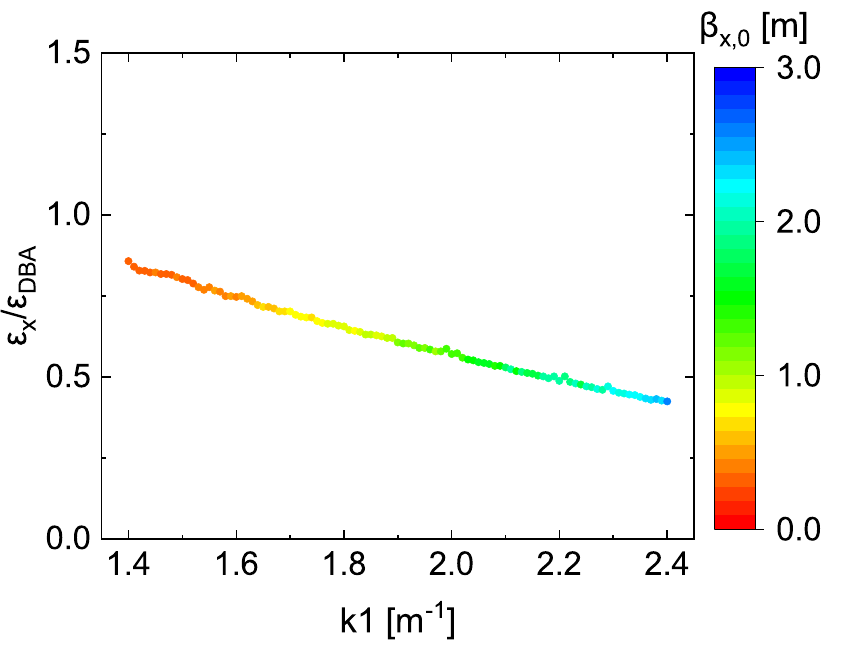}
    \end{minipage}
\caption{\label{fig:f4_SB_DBA_emi} Left: Pareto front of split-bend quadrupole strength $k1$ and emittance ratio $\varepsilon_{x}/\varepsilon_{DBA}$ for a DBA cell model, where the colour denotes the cell horizontal phase advance. Right: $k1$ vs. minimum $\varepsilon_{x}/\varepsilon_{DBA}$ for the DBA cell model with phase advances of about (3$\pi$, $\pi$), where the colour denotes the horizontal beta function at the middle of straight section.}
\end{figure*} 

Inspired by the studies in \cite{ref-15,ref-16-splitbend}, we introduce split bends as the matching bends so as to increase the phase advances ($\Delta\phi_{x}$, $\Delta\phi_{y}$). The split bend is such a bend which is split into two pieces with a quadrupole in between. It increases phase advances by introducing additional focusing. We first use a simplified unit cell model shown in Fig.~\ref{fig:f2_SB_TME} to study the focusing property of split bend. Similar cell model has been mentioned in Ref. \cite{ref-15}, where the split bend cell was studied with fixed phase advances. Thin lens approximation is used for quadrupoles in this model. In the study, for simplicity, the ratio of the bend length to the total drift length is set to 1. Here we set the length of drift $d1$ to 1 m; the half of the split bend, $b1$, has a length of 1 m and a bending angle $\theta$/2 of 1°. The whole ring consists of 180 identical cells. The strengths $k1$, $k2$, $k3$ of three quadrupole families $q1$, $q2$, $q3$ are allowed to vary in the ranges of [0, 5], [-5, 0] and [0, 5] $m^{-1}$, respectively. We scanned the quadrupole strengths with the step of 0.01 $m^{-1}$.

Fig.~\ref{fig:f3_SB_scan_1} presents the results of scanned solutions. The emittances of solutions are compared to the theoretical minimum emittance (TME) \cite{ref-18-TME}, $\varepsilon_{TME}$, with the same bending angle $\theta$, which are shown as ratios of $\varepsilon_{x}/\varepsilon_{TME}$ in the figure. The solutions with emittances larger than 5×$\varepsilon_{TME}$ are not shown in the figure. The split bend cell becomes a TME-like cell when $k1=0$. When $k1$ has a small value, the horizontal phase advance of the unit cell, $\mu_{x}$, is less than 0.5×2$\pi$. As $k1$ increases, $\mu_{x}$ shifts its maximum towards 2$\pi$, and lower emittance can be obtained. This means that we can get a horizontal phase advance of 1.5×2$\pi$ with two split bend unit cells. The red area indicates that the emittance can be reduced to lower than $\varepsilon_{TME}$ with appropriate values of $k1$ and $\mu_{x}$.

\begin{figure}
\includegraphics[width=0.95\linewidth]{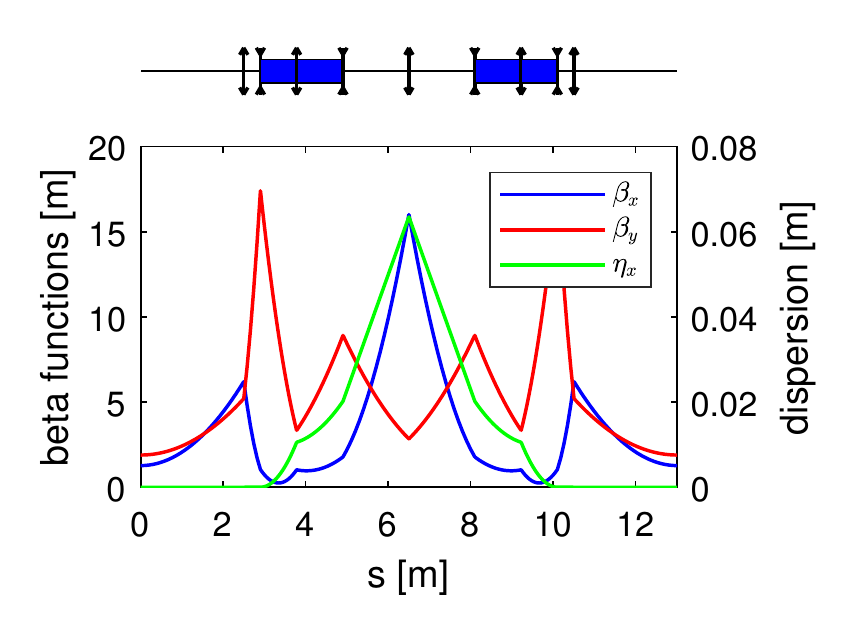}
\caption{\label{fig:f5_SB_DBA_layout} Simplified model layout and optical functions of a DBA lattice cell with split bends, which has horizontal and vertical phase advances of about (3$\pi$, $\pi$). In the layout, split bends are shown in blue and thin-lens quadrupoles are shown as arrows.}
\end{figure}

Our goal is to make an HMBA lattice with –$\it{I}$ across straight section by employing split bends. The –$\it{I}$ part of this HMBA lattice can be seen as a double-bend achromat (DBA) lattice cell with two split bends, which has horizontal and vertical phase advances of about (3$\pi$, $\pi$). Then we study the DBA cell with split bends using a simplified model similar to the unit cell model above. As shown at the top of Fig.~\ref{fig:f5_SB_DBA_layout}, the layout from the first quadrupole to the center of the DBA cell is like the model in Fig.~\ref{fig:f2_SB_TME}. The difference is that the parameters and positions of magnets are not symmetric about the quadrupole inserted in the split bend (hereafter called split-bend quadrupole). In the DBA cell model, the sum of bending angles of two pieces of the split bend is also set to $\theta$=2° and the sum of their lengths is 2 m, and the sum of two different drifts between quadrupoles is 2 m (i.e. here the ratio of the bend length to the total length of two drifts is also 1). Besides, the length of the straight section is set to 5 m. With these settings, the NSGA-III algorithm \cite{ref-19-NSGA-III} was used to study the DBA cell, with magnet parameters and positions as variables.

In the study, the optimization objectives are to minimize the split-bend quadrupole strength $k1$ and the emittance ratio $\varepsilon_{x}/\varepsilon_{DBA}$, where $\varepsilon_{DBA}$ is the minimum emittance of DBA lattice, with $\varepsilon_{DBA}=3\times\varepsilon_{TME}$ \cite{ref-20-TME-Dis}. For the constraints, first we only considered the dispersion-free condition: the absolute dispersion in the straight section is less than 0.001 m. The optimized Pareto front of the two objectives is shown in the left plot of Fig.~\ref{fig:f4_SB_DBA_emi}, and the horizontal phase advance of the DBA cell is also calculated and shown in the figure for these Pareto solutions. We see that as $k1$ increases, the minimum emittance obtained decreases roughly linearly, and the cell horizontal phase advance increases. When $k1=0$, i.e. the usual bend is used, the emittance is about $1.5\times\varepsilon_{DBA}$.

Then, the cell phase advance condition was further considered as the second constraint in the study. For this constraint, we set the horizontal phase advance in the range of 1.48$\sim$1.52×2$\pi$ and the vertical in the range of 0.48$\sim$0.52×2$\pi$. Note that the horizontal and vertical phase advances can not be exactly equal to 3$\pi$ and $\pi$, respectively, for having a stable solution. Due to that the dispersion-free and phase advance constraints we set here are very stringent, it is very difficult to obtain the true Pareto front for the two objectives \cite{ref-21-xu}. So we scanned $k1$ so as to obtain better results, and at each value of $k1$, $\varepsilon_{x}/\varepsilon_{DBA}$ was minimized. The optimization results are shown in the right plot of Fig.~\ref{fig:f4_SB_DBA_emi}, and for these solutions, their horizontal beta functions at the middle of straight section, $\beta_{x,0}$, are also shown as additional information. We see that with the phase advance constraint further considered, the emittance also decreases roughly linearly with increasing $k1$ and the minimum emittance is slightly less than $0.5\times\varepsilon_{DBA}$. We also see that as $k1$ increases, $\beta_{x,0}$ increases but is less than 3 m. The optimum beta function for matching the electron and photon beam phase spaces to enhance the ID brightness is $\beta_{x,0} =L/\pi$ \cite{ref-22-undulator} with $\it{L}$ being the ID length. It is interesting to see that at smaller $k1$, the emittance is larger and $\beta_{x,0}$ is too small (less than 1 m), both of which are not good for brightness enhancement. While at larger $k1$, both lower emittance and reasonably low $\beta_{x,0}$ can promise higher brightness.

Such a DBA lattice cell with phase advances of about (3$\pi$, $\pi$) is shown in Fig.~\ref{fig:f5_SB_DBA_layout}. The emittance of this DBA lattice is 0.64$\times\varepsilon_{DBA}$. The split bend of this DBA lattice is not divided into two identical pieces as that of the unit cell shown in Fig.~\ref{fig:f2_SB_TME}. The outer piece at lower dispersion has a shorter length and higher dipole field compared to the inner piece. As the same principle of longitudinal gradient bend (LGB) \cite{ref-23-LGB1,ref-24-LGB2} for reducing the emittance, distributing high dipole field at the position with low dispersion can help to suppress the quantum excitation, thus giving small dispersion action $\mathcal{H}_{x}$. Besides, the horizontal and vertical beta functions in the straight section have small values so that the ID brightness can be further enhanced.

In addition, the focusing in the split bend of this DBA lattice is relatively strong. When other bends are inserted in the middle part to make the new HMBA lattice, the focusing in these bends can also become strong due to the need for better optics matching, which will be shown in the practical lattice design of the next section.

\section{Examples}
 
As examples, an H8BA lattice and an H6BA lattice with sextupole cancellation across straight section were preliminarily designed. The two designs have the same energy of 3 GeV and the same storage ring circumference of 460.8 m. The H8BA design has 20 lattice cells, and the H6BA design has 24 cells. The NSGA-III algorithm was also used in the lattice design. Following Ref. \cite{ref-7-H10BA}, in the linear optics optimization, not only the natural emittance but also the natural chromaticities and integrated strengths of sextupoles were optimized, so that the lattice solutions with potentially good nonlinear dynamics can be easily got. The main storage ring parameters of the two designs are listed in Table \ref{tab:table1}.

\begin{table}
\caption{\label{tab:table1} Main storage ring parameters of two lattices}

\begin{ruledtabular}
\begin{tabular}{lcc}
\textrm{Parameters}&
\textrm{H8BA}&
\textrm{H6BA}\\
\colrule
Energy [GeV] & \multicolumn{2}{c}{3}\\
Circumference [m] & \multicolumn{2}{c}{460.8}\\
Number of cells & 20 & 24\\
Length of the straight \\ section [m] &  \multicolumn{2}{c}{5} \\
Natural emittance [pm$\cdot$rad] & 74.5 & 136.2\\
Betatron tunes (H/V) & 56.14/15.18 & 52.21/17.13\\
Natural chromaticities \\ (H/V) & -106.6/-100.0 & -94.3/-88.9\\
Momentum compaction & 1.29×$10^{-4}$ & 1.47×$10^{-4}$\\
Damping partition \\ numbers (H/V/L) & 1.91/1.0/1.09 & 1.58/1.0/1.41\\
Natural damping times \\ (H/V/L) [ms] & 13.8/26.3/24.0 & 14.3/22.6/16.0\\
Energy lost per turn [keV] & 351.7 & 407.8\\
Total absolute bending \\ angle [°] & 410.4 & 406.7\\
$\beta_{x}$/$\beta_{y}$ at the middle of \\ straight section [m] & 2.88/1.83 & 2.48/2.39\\
\end{tabular}
\end{ruledtabular}

\end{table}

\begin{figure}
\includegraphics[width=0.95\linewidth]{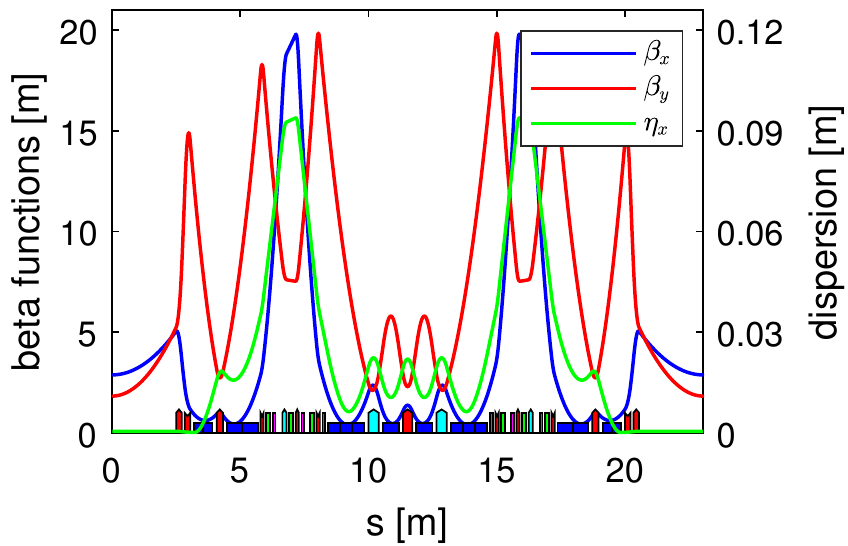}
\caption{\label{fig:f6_SB_H8BA_layout} Magnet layout and optical functions of the H8BA lattice cell. Main bends are in blue, RBs in cyan, quadrupoles in red and sextupoles in green.}
\end{figure}

Fig. \ref{fig:f6_SB_H8BA_layout} shows the magnet layout and linear optics of the designed H8BA lattice cell. Due to that the matching bend, i.e. the split bend, has two pieces of bends, here this lattice is regarded as an H8BA lattice. In this lattice cell, the horizontal and vertical phase advances between the starting position and the peak of the first dispersion bump are about half of (3$\pi$, $\pi$). Except that the two central bends are combined-function bends, the other main bends are LGBs, and two families of RBs are also employed to help reduce the natural emittance down to 74.5 pm·rad. The beta functions at the middle of the straight section have values of 1.5$\sim$3 m. Such small beta functions and the ultra-low emittance can significantly increase the ID brightness.

This H8BA lattice can be compared with the recent H9BA lattice of SSRL-X \cite{ref-13-SSRL-X}, which also uses split bends and has low beta functions in the straight section. But the main arc section of the H9BA lattice is constrained by –$\it{I}$. The horizontal tune of the H9BA lattice cell is about 2.76, while the H8BA lattice cell has a larger horizontal tune of about 2.81, though it has smaller number of bends. This means that the bend unit cells of the H8BA lattice have stronger focusing. For an emittance goal, smaller number of bends together with stronger focusing will generally give higher dipole fields and thus shorter damping times, which is beneficial for suppressing the IBS induced emittance increase. 

\begin{figure}
\centering
\includegraphics[width=0.95\linewidth]{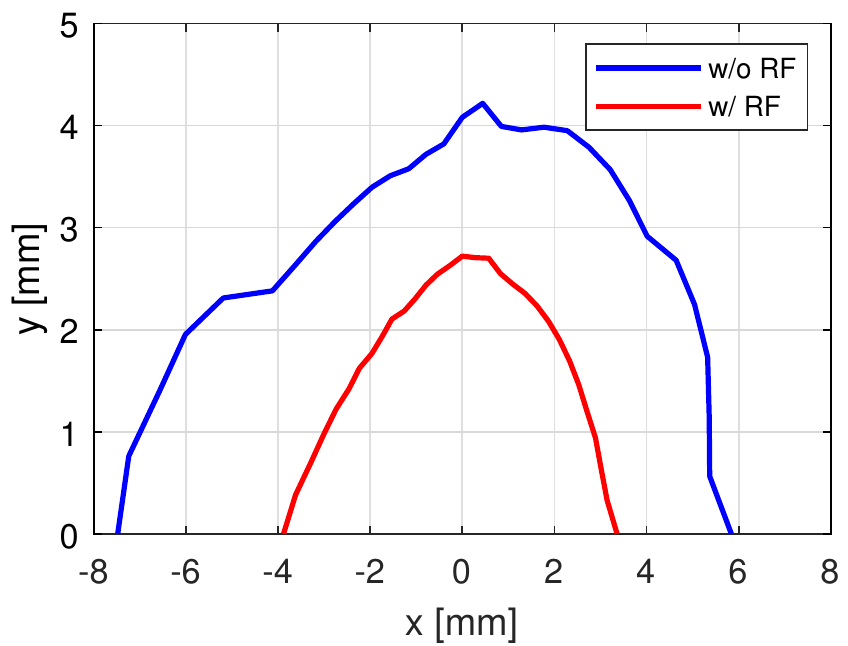}
\caption{\label{fig:f7_H8BA_DA} On-momentum DA of the H8BA lattice, tracked at the middle of the straight section.}
\end{figure}

In the H8BA lattice, three families of sextupoles were used to correct the chromaticities and optimize the nonlinear dynamics. Fig.~\ref{fig:f7_H8BA_DA} shows the optimized on-momentum DA, tracked with ELEGANT \cite{ref-25-Elegant}. We see that the horizontal 4D DA is about -8$\sim$6 mm, which is relatively large considering the ultra-low emittance and the small horizontal beta function. With RF cavity included, the DA shrinks due to the path lengthening effect as in other HMBA lattices.

\begin{figure}
    \begin{minipage}[t]{0.48\linewidth}
		\centering
		\includegraphics{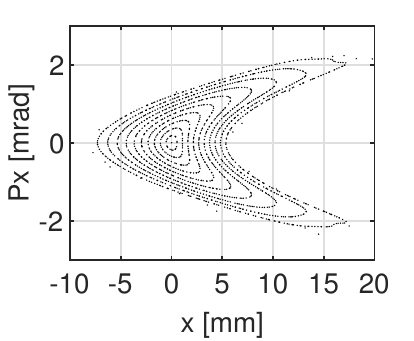}
		\includegraphics{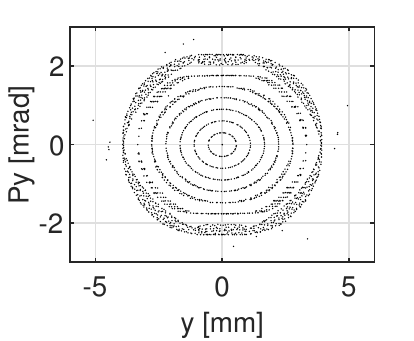}
    \end{minipage}
    \begin{minipage}[t]{0.48\linewidth}
	    \centering
	      \includegraphics{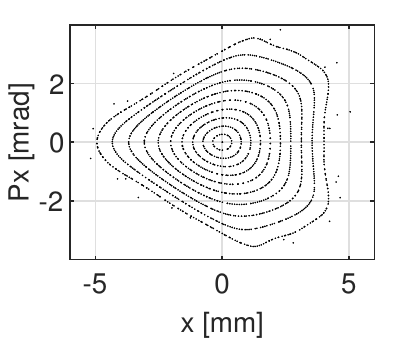}
            \includegraphics{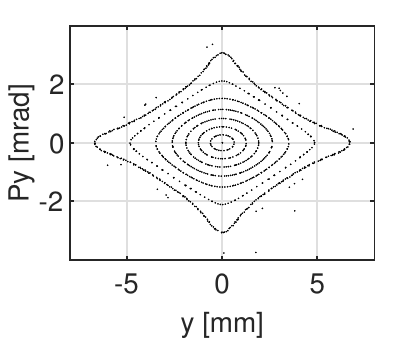}
    \end{minipage}
\caption{\label{fig:f8_Phasespace} Horizontal (top) and vertical (bottom) phase space tori of the H8BA lattice, tracked at the middle of the straight section (left) and the middle of the arc section (right). }
\end{figure}

Due to the –$\it{I}$ made across the straight section, the sextupole resonance driving terms \cite{ref-26-RDT} are not cancelled when observed in the straight section, and they are cancelled when observed in the middle part of the arc section. Fig.~\ref{fig:f8_Phasespace} shows the transverse phase space tori tracked at two locations. We see that the horizontal phase space tori are much distorted at the middle of the straight section. Besides, IDs installed in straight sections may disturb the nonlinear cancellation, while the low beta functions in straight sections are beneficial for reducing the tune shifts and beta-beat produced by IDs. Here we will not further study the ID effect and compensation. Anyway, for this lattice, we can adopt the on-axis swap-out injection \cite{ref-27-swap-out-injection}, which can allow a DA of only $\sim$1 mm.

\begin{figure}
\includegraphics[width=0.95\linewidth]{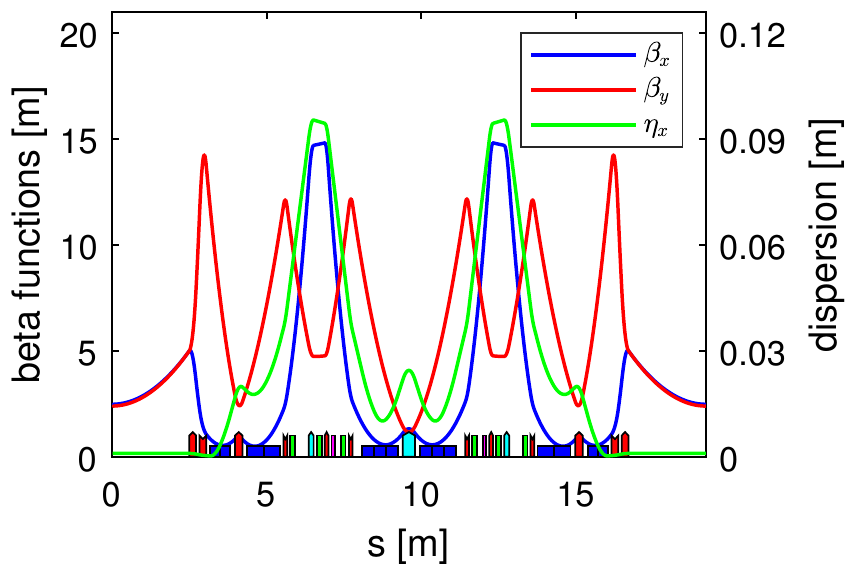}
\caption{\label{fig:f9_SB_H6BA_layout} Magnet layout and optical functions of the H6BA lattice cell. Octupoles are in pink, and the other magnets are as in Fig. \ref{fig:f6_SB_H8BA_layout}.}
\end{figure}

To reflect the flexibility of the new lattice concept, an H6BA lattice is also designed, shown in Fig. \ref{fig:f9_SB_H6BA_layout}, which removes the two central combined-function bends from the H8BA lattice. The natural emittance is 136.2 pm·rad. Recently, PETRA IV also adopted an H6BA lattice \cite{ref-8-PETRA-H6BA} with split bends used and low beta in the straight, but the two central bend cells are constrained by –$\it{I}$ with transverse phase advances of ($\pi$, $\pi$). The horizontal tune of the H6BA lattice cell here is near 2.2, obviously larger than the value of less than 1.8 for the PETRA IV H6BA lattice cell, which is beneficial for reducing the emittance. As said before, this can also reduce the damping times for an emittance goal so as to better suppress the IBS effect. It is interesting to note that the central two bends can also be seen as a split bend. If we regard the split bend as a single bend, then this H6BA lattice can be seen as a triple-bend achromat lattice.

\begin{figure}
\includegraphics[width=0.95\linewidth]{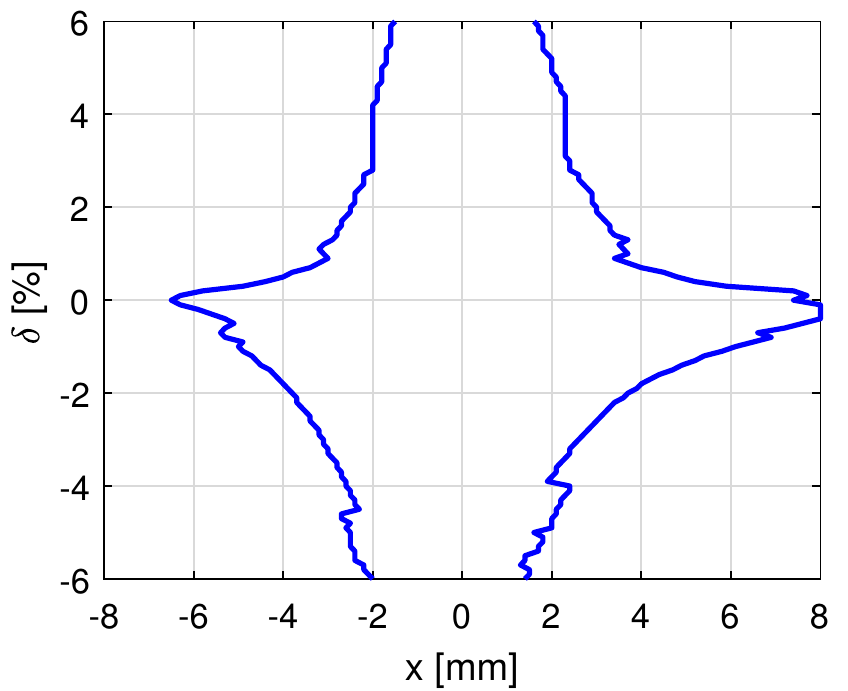}
\caption{\label{fig:f10_H6BA_offset_DA} Horizontal DAs of the H6BA lattice for relative momentum deviations of -6\%$\sim$6\%, tracked at the middle of the straight section.}
\end{figure}

\begin{figure}
\includegraphics[width=0.95\linewidth]{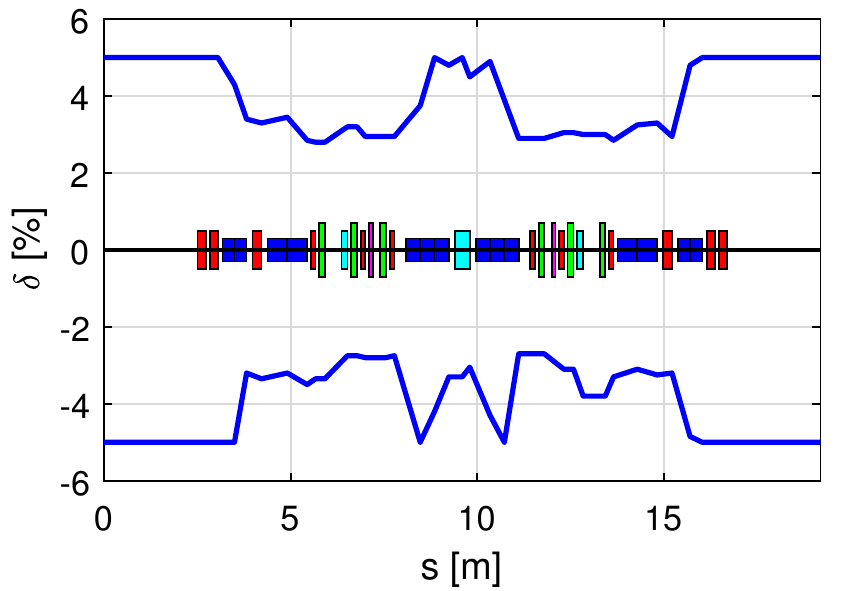}
\caption{\label{fig:f11_H6BA_LMA} Local momentum aperture along one H6BA lattice cell. The RF momentum acceptance is set to about 5\%.}
\end{figure}

Fig.~\ref{fig:f10_H6BA_offset_DA} shows the horizontal DAs for a range of momentum deviations, optimized with three sextupole families and one octupole family. We see that the on-momentum horizontal DA is larger than 6 mm and the dynamic momentum acceptance is large. The local momentum aperture (LMA) is shown in Fig.~\ref{fig:f11_H6BA_LMA}. Based on this error-free LMA, the Touschek lifetime was calculated, which is about 9.7 hours for 10\%-coupling beam with bunch charge of 1 nC. In the calculation, the frequency of RF cavity is 500 MHz, and the bunch is assumed to be lengthened by a factor of 4 with harmonic cavity. 

The H6BA design has more straight sections than the H8BA, and we can place some DWs in straight sections to further reduce the emittance and also to reduce the damping times for suppressing IBS induced emittance increase. With 4 DWs employed, each with length of 4.2 m, peak field of 2.2 T and period of 105 mm, the natural emittance is reduced to 88.4 pm·rad. And the damping times (H/V/L) are reduced to 8.3/10.6/6.1 ms with $J_x$ of 1.27, smaller than that without DW. When including IBS, the horizontal emittance increases to 99.2 pm·rad for the beam as in the Touschek lifetime calculation. If full-coupling beam is used, the horizontal emittance can be further reduced and the horizontal emittance increase due to IBS can also be better suppressed.

\section{Extension}

\begin{figure}
\includegraphics[width=0.95\linewidth]{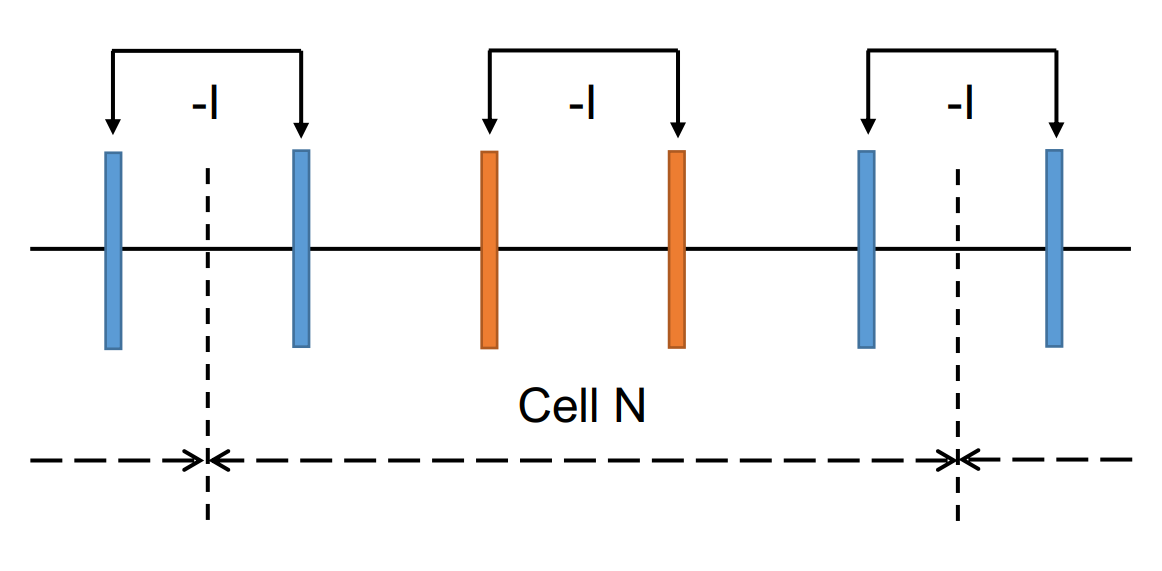}
\caption{\label{fig:f12_extension} Schematic of an MBA lattice with two pairs of –$\it{I}$ separated dispersion bumps.}
\end{figure}

The nonlinear cancellation scheme proposed above can be extended to produce other MBA lattices. If the emittances of DLSRs are further reduced, such as the $\sim$10 pm·rad emittance design for MAX IV future upgrade \cite{ref-28-MAXIV-upgrade}, the number of bends in an MBA lattice will increase. If conventional MBA lattices are used, like a 19BA lattice for MAX IV upgrade \cite{ref-28-MAXIV-upgrade}, the magnet strengths will be extremely strong and the magnet layout will be unusually compact. A possible solution to mitigate the magnet and space issue is to use an HMBA-like lattice with few dispersion bumps. If there are only one pair of bumps in an MBA lattice with many bends, like the ALS-U 9BA lattice \cite{ref-29-ALS-U-9BA} and the H10BA lattice \cite{ref-7-H10BA}, it is very hard to have good off-momentum nonlinear dynamics for reasonable beam lifetime. Introducing another pair of bumps may improve the off-momentum nonlinear dynamics \cite{ref-30-IDB-MBA}. Fig.~\ref{fig:f12_extension} shows the schematic of an MBA lattice with two pairs of –$\it{I}$ separated bumps. The two pairs of bumps are non-interleaved: the inner pair is inside the cell and the outer pair is the same as the scheme proposed in this paper. This non-interleaved layout is good for improving nonlinear dynamics performance.

Besides, due to that the inner part of the arc section in the new HMBA lattice is free, it can be easily extended to a combined lattice, H$m$BA-H$n$BA lattice ($m$$\neq$$n$), without affecting the nonlinear cancellation.

\section{Conclusion}

In this paper, we studied a new HMBA lattice concept to make the main arc section free of the phase advance constraint so as to increase the flexibility of lattice. In the concept, –$\it{I}$ transformation for sextupole cancellation is moved to both sides of the lattice cell. To increase the phase advances for satisfying the –$\it{I}$, split bends are used as matching bends, which also reduce the emittance, and beta functions in the straight section are lowered, which is also good for enhancing ID brightness. The bend unit cells of the new HMBA lattice can have stronger focusing than those of the typical HMBA lattice, which is beneficial for reducing the emittance and also for suppressing the IBS effect for a given emittance goal. The DA of the new lattice is relatively large, but the phase space tori are distorted in the straight section due to that the nonlinear cancellation is made across the straight section. This cancellation can also be extended to produce other MBA lattices for DLSRs.

\begin{acknowledgments}
This work was supported by the National Key Research and Development Program of China under Grant No. 2016YFA0402000 and the National Natural Science Foundation of China under Grant Nos. 11875259 and 12205299.
\end{acknowledgments}

\bibliography{ref.bib}

\providecommand{\noopsort}[1]{}\providecommand{\singleletter}[1]{#1}%
\begin{thebibliography}{10}

\bibitem{ref-1-MBA}
D.~Einfeld et~al.
\newblock {First multi-bend achromat lattice consideration}.
\newblock {\em {Journal of Synchrotron Radiation}}, 21(5):856--861, 2014.

\bibitem{ref-2-HMBA}
L.~Farvacque et~al.
\newblock {A Low-Emittance Lattice for the ESRF}.
\newblock In {\em Proc. IPAC'13}, pages 79--81. JACoW Publishing, Geneva,
  Switzerland.

\bibitem{ref-3-APSU}
M.~Borland et~al.
\newblock {Lower Emittance Lattice for the Advanced Photon Source Upgrade Using
  Reverse Bending Magnets}.
\newblock In {\em Proc. NAPAC'16}, pages 877--880. JACoW Publishing, Geneva,
  Switzerland.

\bibitem{ref-4-antibend}
A.~Streun.
\newblock {The anti-bend cell for ultralow emittance storage ring lattices}.
\newblock {\em Nuclear Instruments and Methods in Physics Research Section A:
  Accelerators, Spectrometers, Detectors and Associated Equipment},
  737:148--154, 2014.

\bibitem{ref-5-Diamond}
A.~Alekou et~al.
\newblock {Study of a Double Triple Bend Achromat (DTBA) Lattice for a 3 GeV
  Light Source}.
\newblock In {\em Proc. IPAC'16}, pages 2940--2942. JACoW Publishing, Geneva,
  Switzerland.

\bibitem{ref-6-HALF}
Z.~H. Bai et~al.
\newblock {A Modified Hybrid 6BA Lattice for the HALF Storage Ring}.
\newblock In {\em Proc. IPAC'21}, pages 407--409. JACoW Publishing, Geneva,
  Switzerland.

\bibitem{ref-7-H10BA}
P.~H. Yang et~al.
\newblock {Design of a hybrid ten-bend-achromat lattice for a
  diffraction-limited storage ring light source}.
\newblock {\em Nuclear Instruments and Methods in Physics Research Section A:
  Accelerators, Spectrometers, Detectors and Associated Equipment}, 943:162506,
  2019.

\bibitem{ref-8-PETRA-H6BA}
I.~V. Agapov et~al.
\newblock {PETRA IV Storage Ring Design}.
\newblock In {\em Proc. IPAC'22}, number~13 in International Particle
  Accelerator Conference, pages 1431--1434. JACoW Publishing, Geneva,
  Switzerland, 7 2022.

\bibitem{ref-9-MBA-lattice1}
Andreas Streun et~al.
\newblock {SLS-2--the upgrade of the Swiss Light Source}.
\newblock {\em Journal of synchrotron radiation}, 25(3):631--641, 2018.

\bibitem{ref-10-MBA-lattice2}
Penghui Yang et~al.
\newblock {Design of a diffraction-limited storage ring lattice using
  longitudinal gradient bends and reverse bends}.
\newblock {\em Nuclear Instruments and Methods in Physics Research Section A:
  Accelerators, Spectrometers, Detectors and Associated Equipment}, 990:164968,
  2021.

\bibitem{ref-11-MBA-lattice3}
G~Baranov et~al.
\newblock {Lattice optimization of a fourth-generation synchrotron radiation
  light source in Novosibirsk}.
\newblock {\em Physical Review Accelerators and Beams}, 24(12):120704, 2021.

\bibitem{ref-12}
Z.~H. Bai.
\newblock {Lattice design progress of the HALF storage ring, LEL 2022 - 3rd
  Workshop on Low Emittance Lattice Design}, 2022.

\bibitem{ref-13-SSRL-X}
J.~Kim et~al.
\newblock {A Hybrid Multi-Bend Achromat Lattice Design for SSRL-X}.
\newblock In {\em Proc. IPAC'22}, number~13 in International Particle
  Accelerator Conference, pages 207--209. JACoW Publishing, Geneva,
  Switzerland, 7 2022.

\bibitem{ref-14-korean}
G.~S. Jang et~al.
\newblock {Low emittance lattice design for Korea-4GSR}.
\newblock {\em Nuclear Instruments and Methods in Physics Research Section A:
  Accelerators, Spectrometers, Detectors and Associated Equipment},
  1034:166779, 2022.

\bibitem{ref-15}
Anton Bogomyagkov and other.
\newblock {Low emittance lattice cell with large dynamic aperture}.
\newblock {\em arXiv preprint arXiv:1405.7501}, 2014.

\bibitem{ref-16-splitbend}
P.~Raimondi.
\newblock {Beyond EBS, ESRF science Live Online Seminars}, 2021, 04 2021.

\bibitem{ref-17-SOLEIL}
A.~Loulergue et~al.
\newblock {Baseline Lattice for the Upgrade of SOLEIL}.
\newblock In {\em Proc. IPAC'18}, pages 4726--4729. JACoW Publishing, Geneva,
  Switzerland.

\bibitem{ref-18-TME}
L.C. Teng.
\newblock {Minimizing the emittance in designing the lattice of an electron
  storage ring}.
\newblock Technical Report TM-1269, Fermi National Accelerator Lab., 1984.

\bibitem{ref-19-NSGA-III}
Yuan Yuan, Hua Xu, and Bo~Wang.
\newblock An improved nsga-iii procedure for evolutionary many-objective
  optimization.
\newblock GECCO '14, page 661–668, New York, NY, USA, 2014. Association for
  Computing Machinery.

\bibitem{ref-20-TME-Dis}
S.Y. Lee and L.~Teng.
\newblock {Theoretical minimum emittance lattice for an electron storage ring}.
\newblock In {\em Conference Record of the 1991 IEEE Particle Accelerator
  Conference}, pages 2679--2681 vol.5, 1991.

\bibitem{ref-21-xu}
J.~H. Xu et~al.
\newblock {Constraint handling in constrained optimization of a storage ring
  multi-bend-achromat lattice}.
\newblock {\em Nuclear Instruments and Methods in Physics Research Section A:
  Accelerators, Spectrometers, Detectors and Associated Equipment}, 988:164890,
  2021.

\bibitem{ref-22-undulator}
Ryan~R. Lindberg and Kwang-Je Kim.
\newblock {Compact representations of partially coherent undulator radiation
  suitable for wave propagation}.
\newblock {\em Phys. Rev. ST Accel. Beams}, 18:090702, Sep 2015.

\bibitem{ref-23-LGB1}
Ryutaro Nagaoka and Albin~F Wrulich.
\newblock {Emittance minimisation with longitudinal dipole field variation}.
\newblock {\em Nuclear Instruments and Methods in Physics Research Section A:
  Accelerators, Spectrometers, Detectors and Associated Equipment},
  575(3):292--304, 2007.

\bibitem{ref-24-LGB2}
Andreas Streun and Albin Wrulich.
\newblock {Compact low emittance light sources based on longitudinal gradient
  bending magnets}.
\newblock {\em Nuclear Instruments and Methods in Physics Research Section A:
  Accelerators, Spectrometers, Detectors and Associated Equipment},
  770:98--112, 2015.

\bibitem{ref-25-Elegant}
M~Borland.
\newblock Elegant: A flexible sdds-compliant code for accelerator simulation.
\newblock 8 2000.

\bibitem{ref-26-RDT}
J.~Bengtsson.
\newblock {\em {The Sextupole Scheme for the Swiss Light Source (SLS): An
  Analytic Approach, 1997}}, 1997.

\bibitem{ref-27-swap-out-injection}
L.~Emery and M.~Borland.
\newblock {Possible Long-Term Improvements to the Advanced Photon Source}.
\newblock In {\em Proc. PAC'03}, pages 256--258. JACoW Publishing, Geneva,
  Switzerland.

\bibitem{ref-28-MAXIV-upgrade}
Pedro~Fernandes Tavares et~al.
\newblock {Future development plans for the MAX IV light source: Pushing
  further towards higher brightness and coherence}.
\newblock {\em Journal of Electron Spectroscopy and Related Phenomena},
  224:8--16, 2018.

\bibitem{ref-29-ALS-U-9BA}
C.~Sun et~al.
\newblock {Design of the ALS-U Storage Ring Lattice}.
\newblock In {\em Proc. IPAC'17}, pages 2827--2829. JACoW Publishing, Geneva,
  Switzerland.

\bibitem{ref-30-IDB-MBA}
Z.~H. Bai et~al.
\newblock {Study of Seven-Bend Achromat Lattices with Interleaved Dispersion
  Bumps for HALS}.
\newblock In {\em Proc. IPAC'19}, pages 1495--1497. JACoW Publishing, Geneva,
  Switzerland.

\end{thebibliography}

\end{document}